\documentclass[]{aa}
\usepackage{amsmath}
\usepackage{graphicx,color}
\usepackage{natbib}
\usepackage{microtype}
\usepackage{url}
\usepackage[varg]{txfonts}
\usepackage{threeparttable}
\usepackage{booktabs, makecell}
\usepackage{longtable}
\usepackage[utf8]{inputenc}
\usepackage{amsmath}
\usepackage{verbatim}
\usepackage{txfonts}

\usepackage[colorlinks=true,allcolors=blue,draft=true]{hyperref}
\def\gai{\textit{Gaia}\xspace}
\def\ztf{\textit{ZTF}\xspace}
\def\glx{\textit{GALEX}\xspace}

\def\ero{eROSITA\xspace}
\def\xmmn{\textit{XMM-Newton}\xspace}

\def\xmm{{XMM152737}\xspace}
\def\pnstr{\textit{Pan-STARRS}\xspace}

\newcommand\fergs{\ensuremath{\mathrm{erg}\,\mathrm{cm}^{-2}\,\mathrm{s}^{-1}}\xspace}
\newcommand\lergs{\ensuremath{\mathrm{erg}\,\mathrm{s}^{-1}}\xspace}

\begin{document} 

\title{Discovery of the magnetic cataclysmic variable XMM~J152737.4-205305.9 with a deep eclipse-like feature}

\author{Samet Ok \inst{1,2}
           \and Axel Schwope \inst{1}
            \and David A. H. Buckley \inst{3,4,5}
            \and Jaco Brink \inst{3,4} 
          }
\institute{Leibniz-Institut für Astrophysik Potsdam (AIP), An der Sternwarte 16, 14482 Potsdam, Germany\\
\email{sok@aip.de}
\and
Department of Astronomy \& Space Sciences, Faculty of Science, University of Ege, 35100 Bornova, Izmir, Turkey
\and
South African Astronomical Observatory, PO Box 9, Observatory Road, Observatory 7935, Cape Town, South Africa
\and
Department of Astronomy, University of Cape Town, Private Bag X3, Rondebosch 7701, South Africa
\and
Department of Physics, University of the Free State, PO Box 339, Bloemfontein 9300, South Africa
}

\date{Received ...; Accepted ...}
                
\abstract{In this study, we report a discovery from \xmmn, which involves the identification and subsequent examination of a newly discovered polar-type cataclysmic variable named XMM J152737.4-205305.9. The discovery was made by matching the \xmmn data archive with the cataclysmic variable candidate catalog provided by \gai Data Release 3. The utilization of X-ray photometry has led to the identification of two distinct dips that exhibit a recurring pattern with a precise period of 112.4(1) minutes in two \xmmn observations that are one year apart. The data obtained from the photometry of Zwicky Transient Facility (ZTF) and ATLAS surveys consistently indicate the presence of the different mass accretion states of up to 2 mag. Following the optical data, the \textit{SRG}(Spectrum Roentgen Gamma)/\ero All Sky Survey observed the system in two different X-ray levels which may imply different accretion states. Following these observations, the low-resolution spectrum obtained using SALT spectroscopy exposes the prominent hydrogen Balmer and helium emission lines, strongly supporting that the system belongs to the category of polar-type magnetic cataclysmic variable. The \xmmn observations, conducted under various conditions of X-ray levels, reveal a consistent pattern of a deep dip-like feature with a width of $\approx 9.1$ min. This feature implies the presence of an eclipse in both observations. According to \gai data, the object is located at a distance of $1156^{+720}_{-339}$\,pc, and its X-ray luminosity lies within the $L_{\rm X}$= (3-6)$\times10^{31}$ \lergs range.}

\maketitle
%

\section{Introduction}

Cataclysmic variables (CVs) refer to binary systems characterized by a white dwarf (WD) primary star and a main sequence low mass star as a secondary that is in a semi-detached form, whereby it is actively accreting mass from a secondary \citep{warner+95}. If the magnetic field of the white dwarf (WD) is low or absent, accretion onto the WD takes place through an accretion disc surrounding it. If the WD possesses a magnetic field of moderate or strong intensity (> 10$^{7}$ G), the formation of the disc will either be limited or completely inhibited. Instead, the mass will accrete through the magnetic field lines that fall onto the magnetic poles of the WD.

The systems known as polars (or AM Her systems) exhibit the highest levels of magnetic field strength among the CVs. These systems, characterized by relatively short orbital periods (< 2 $h$), are commonly denoted as soft X-ray emitters \citep{ritter+03,kuulkers+10,mukai17}. 
Strong magnetic field and close orbital interactions keep both stars in synchronous rotation \citep{cropper90}. Due to the locked orbits and the absence of a surrounding disc, these objects host mass accretion in a comparatively more sterile environment than other sub-types of CVs. Hence, it can be inferred that these objects hold significant potential for effectively monitoring mass accretion processes, advancing our understanding of the intricate interplay between plasma and magnetic field interactions.

The occurrence of mass accretion induces the X-ray emissions on the surface of the white dwarf due to the creation of shocks. Hence, it can be observed that these kinds of objects exhibit more prominence within the X-ray region as compared to other wavelengths. In particular, the ROSAT All Sky Survey has been able to increase the numbers of polars considerably \citep{beuermann+99}. 

Polars have a discernable X-ray spectrum that has a slightly flat plateau-like shape, particularly within the X-ray wavelength range of 0.05 to 10 keV \citep{kuulkers+10}. The spectrum obtained from a multi-temperature area is likely to be the result of a mixture of several emission processes and emission regions. One notable component among these is blackbody radiation in soft X-rays, which may be considered particularly distinctive in the polars. The projecting feature observed in the soft X-ray spectrum, which arises from the reprocessed radiation by the white dwarf, is one of the distinctive X-ray features \citep{king+79, lamb+79}. Interestingly, the polar survey conducted by \cite{ramsay+04} with the X-ray Multi-Mirror Mission (\xmmn) and all serendipitous discoveries made with \xmmn since then \citep{vogel+08, ramsay+09, webb+18, schwope+20} did not reveal any further evidence for soft X-ray emission from a large number of magnetic CVs. Also, the first \ero-discovered polar did not show pronounced soft X-ray emission \citep{schwope+22}.

An essential characteristic of polars, which is frequently employed for classification purposes, is the presence of distinctive features in their optical spectrum. One of the most important features that distinguish these objects from other CVs is the emission lines seen in the spectrum. The spectrum is primarily characterized by the H Balmer, with a notable He I (4472 $\AA$ and 5876 \AA) and He II (4686 \AA) emission lines, which is the determining factor in their categorization \citep{szkody+02, breedt+14, thorstensen+16}. These lines often exhibit a single-peaked shape due to the lack of an accretion disk but show asymmetric profiles due to the presence of an accretion stream \citep{schwope+97}

As active mass accretion systems, polars display random brightness changes, since interruptions or limitations in accretion flow might occur occasionally. The observed variations in brightness, which are thought to be associated with the activity of the secondary star \citep{livio+94, kafka+05}, can render the polars rather dim when there is little mass accretion, making them challenging to detect due to their location beyond the detectable range of existing observational instruments. Through the utilization of predominantly long-term sky surveying initiatives, it became feasible to identify those bodies throughout their phase of increased mass accretion. Accordingly, the discovery of many CVs has been made serendipitously with all-sky surveys or pointed observations of the relevant sky region. Here we report a CV discovery obtained in this manner. XMM~J152737.4-205305.9, or hereafter \xmm, was discovered by correlating the CV candidate catalog released by \gai Data Release 3 with the \xmmn archive. We describe the observations that led to our discovery in Sec. 2. In Sec. 3, we present our primary results and discuss some fundamental parameters of the object.

\begin{figure}
\resizebox{\hsize}{!}{\includegraphics[width=\columnwidth]{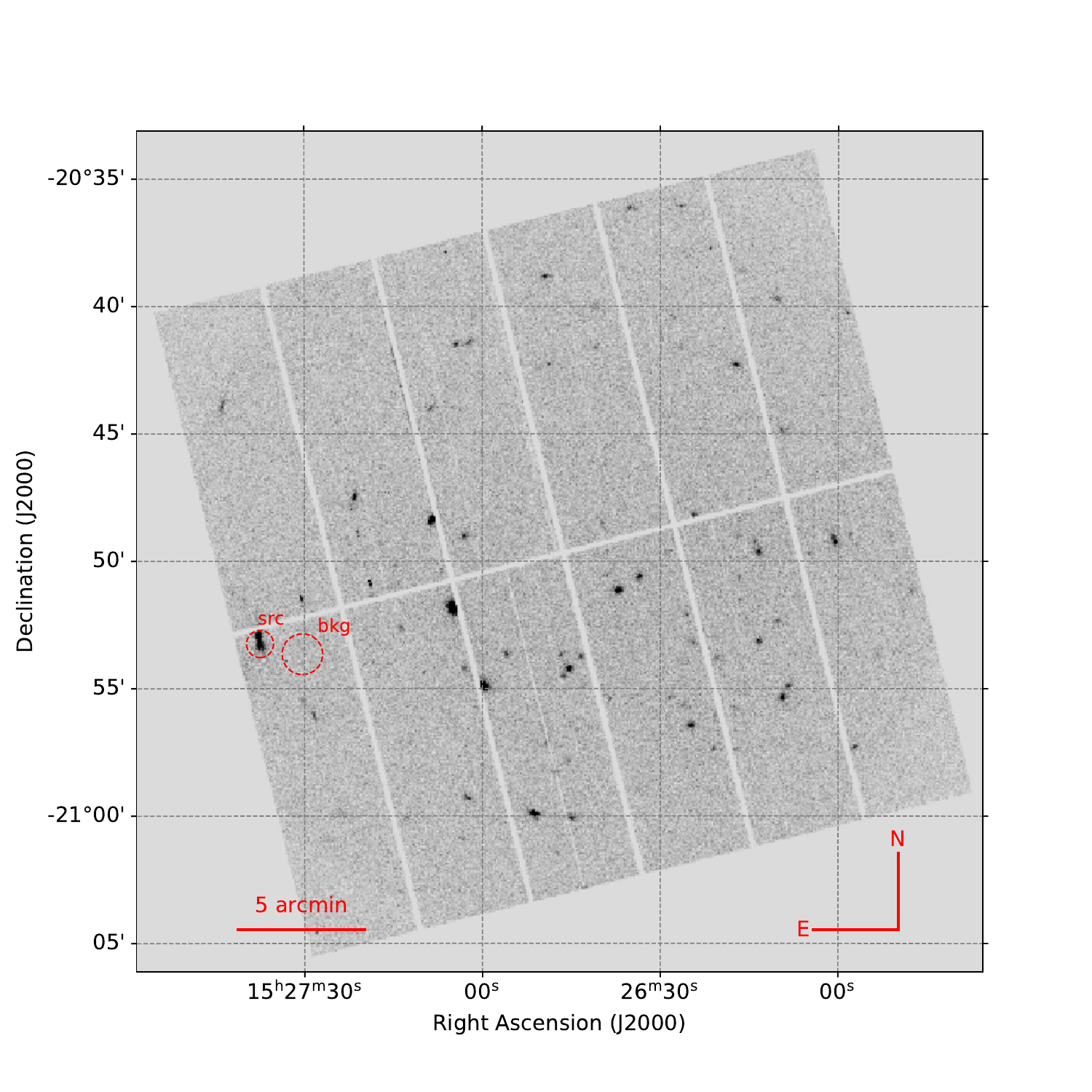}}
\caption{X-ray image of the 0884991701 obs ID of \xmmn observation. The red dashed circle marks the exact position of the \xmm (src) and selected background region (bkg). This image was made using photon energies between $0.2 - 10.0$ keV.
\label{f:xmmsky}}
\end{figure}

\begin{figure}
\resizebox{\hsize}{!}{\includegraphics[width=\columnwidth]{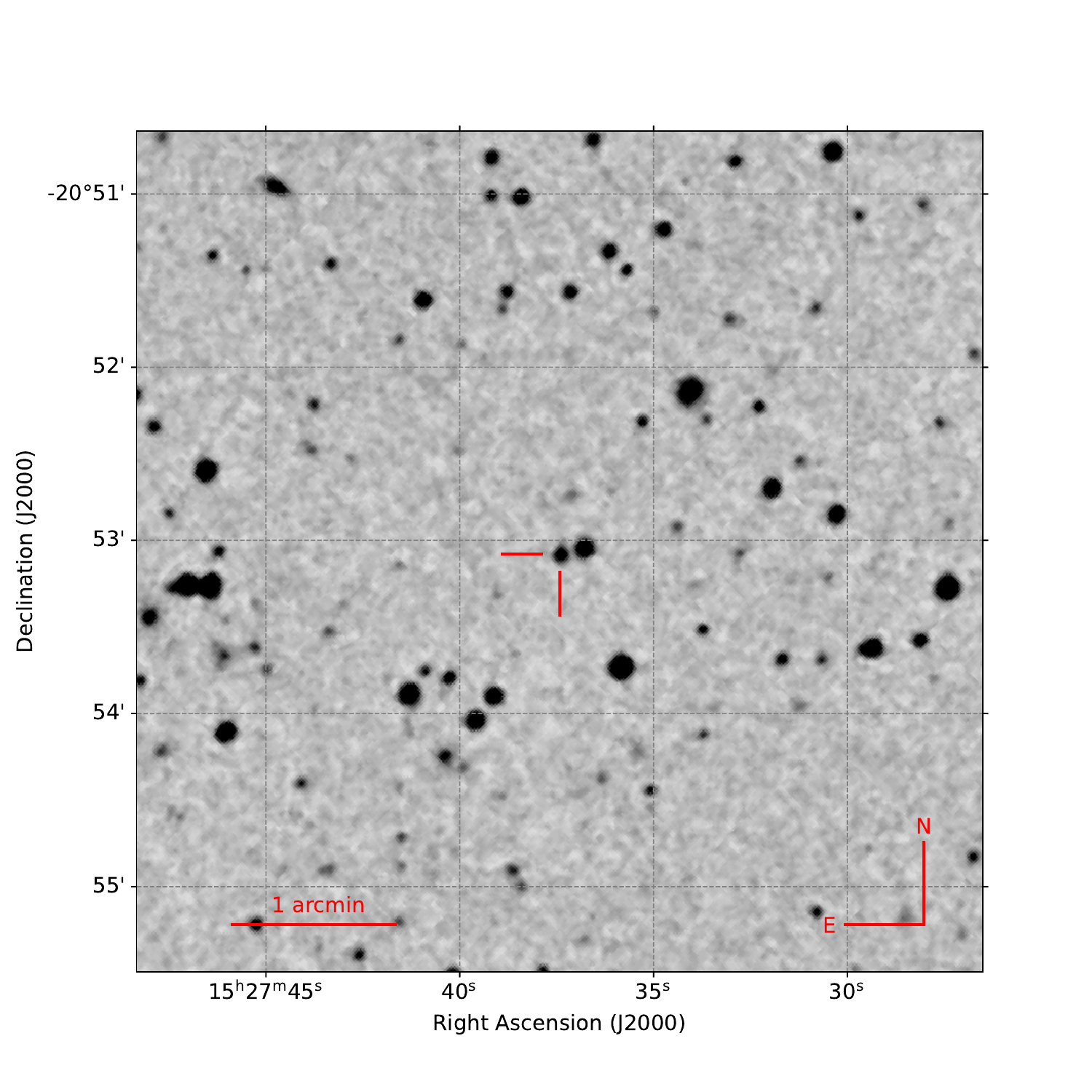}}
\caption{Finding chart for \xmm with North at top and East to the left. The image is from the Digitized Sky Survey (DSS) and the new  object is marked with red lines. A scale bar is shown in the lower left.
\label{f:skymap}}
\end{figure}

\section{Observations \label{s:obs}}

With \gai DR3 \citep{gaia+21}, a catalog for the classification of 12.4 million variable sources has been released. In this catalog, 7306 sources identified based on long-term photometric variations are presented as cataclysmic variable candidates \citep{rimoldini+23}. In order to detect and identify new magnetic cataclysmic variables, we correlated the source coordinates in this catalog within the XMM-Newton's data archive. The matching process is applied with a default search radius predefined by \xmmn database as 18 arcmin for every object. As a result of this search, nearly 400 objects were found to be in at least one observational data set in the \xmmn archive. The following deep research of databases and the literature revealed that the majority of these 400 objects, especially the bright ones, were known cataclysmic variables. We saw remarkably sudden and prominent fluctuations in one of the systems whose light curve we investigated. Deep research has revealed the fast and short-term eclipse-like variations in the \xmmn's X-ray light curve of the \xmm that had never been investigated before.

In order to classify and fully understand the nature of the object, and determine its distinguishing characteristics, we analyzed data from available public archives and performed descriptive low-resolution optical spectroscopy.

\begin{figure}
   \includegraphics[width=\columnwidth]{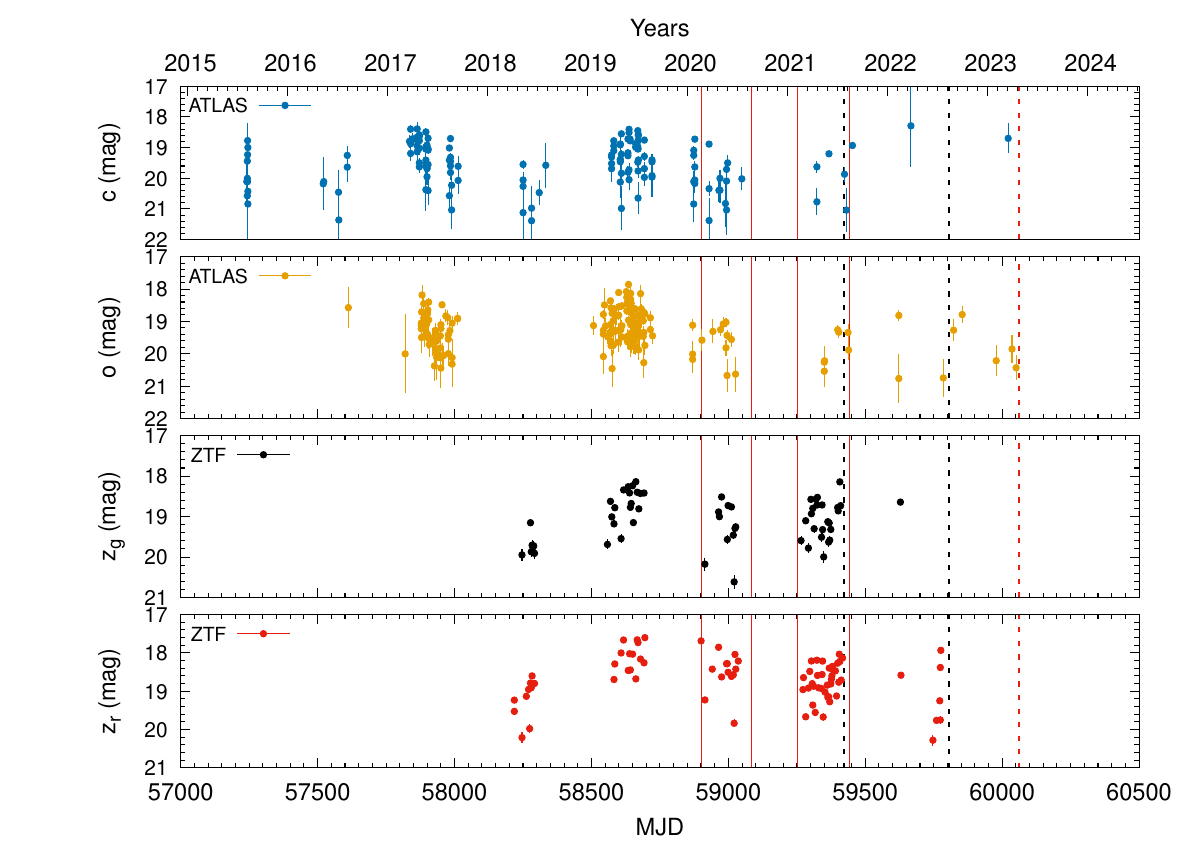}
      \caption{Long-term optical light curve of \xmm which corresponds to ATLAS and ZTF observations. The vertical black dashed lines show when the \xmmn observation occurred. The red dashed line shows the SALT observation. The vertical solid red lines show eROSITA All Sky Survey observations obtained at 6-month intervals.
      } 
      \label{f:atlasztf}
\end{figure}

\subsection{\xmmn observations}

The X-ray observations of the object were performed with \xmmn on July 29, 2021, and August 14, 2022. The pointed observations were carried out under the Multi-Year Heritage program. These observations were planned to observe the quasar QSO J1526-2050. The observation IDs are 0884991701 (Obs 1) and 0886210801 (Obs 2) and were lasting 116 ks and 103 ks, respectively. Both observations were operated in full-frame mode with thin filters. All \xmmn instruments (EPIC-pn, EPIC-MOS, Optical Monitor, and RGS) were used together but due to the smaller fields of view of all other instruments, only EPIC-pn with its large field of view gave useful data \citep{strueder+01}. The source coordinates are given in the \xmmn database as RA=15:27:37.50 and DEC=$-$20:53:05.4 with 0.3 arcsec position errors. The object was discovered at the edge of the detector; an X-ray image of the object obtained during Obs 1 is shown in Fig.~\ref{f:xmmsky}. The system is also noted in the \xmmn's fourth serendipitous source catalog with a source id of 4XMM J152737.4-205305 \citep{webb+20}. 

The data were reduced with Science Analysis System (SAS) software that has been purposely produced to \xmmn data \citep{gabriel17}. The EPIC-pn data were processed with the epchain task to generate calibrated event lists. The time columns in the event list were corrected to the Solar System barycenter using the barycen task and photon lists were filtered to only include photons with energies between 0.2 keV and 10.0 keV. 

There was a severe flare event contaminating X-ray photons at the beginning of Obs 1, lasting approximately 3 hours, between JD=59425.30 and JD=59425.56, and was identified from a high background level. A clean event file was created by filtering the light curve accepting as good times only those where the background rate was $<0.4$ counts s$^{-1}$ by using the tabgtigen task in \textit{SAS}. In Obs 2, this flare activity was observed at random intervals throughout the entire observation. Likewise, in this observation, a clean event file was obtained with tabgtigen.

\subsection{\gai observations}

In \textit{Gaia} DR3 \citep{gaia+21} \xmm is quoted by ID 5290647986316685824 with the sky coordinates of DEC2000=231.906207 deg and RA2000=-20.8852306 deg. The mean brightness of the object is 19.08$\pm0.02$, 19.39$\pm0.06$, and 18.43$\pm0.05$ in the \textit{G}, \textit{$G_{BP}$}, and \textit{$G_{RP}$} passbands, respectively. \textit{Gaia} measured the parallax of \xmm as 1.07$\pm$0.34 mas. We used the geometric distance ($r_{\rm geo}$) to the system of $1156^{+720}_{-339}$\,pc, as determined by \cite{bailer-jones+21}. The object is quite faint, so in addition to \cite{bailer-jones+21}, we also present here the $934^{+436}_{-225}$\,pc distance from the inverse parallax with a 1 sigma error. The sky position of the source is shown in Fig.~\ref{f:skymap}.

\subsection{\textit{ATLAS} observations}

Asteroid Terrestrial-impact Last Alert System (ATLAS) \citep{tonry+18} reports many observations of the object between the years 2015 and 2023. In this interval, there are in total 322 data points in passbands \textit{o} (5582 - 8249 \AA) and \textit{c} (4157 - 6556 \AA). The data were obtained from the interactive \textit{ATLAS}\footnote{\protect\url{https://fallingstar-data.com/forcedphot}} database by using forced photometry. Fig.~\ref{f:atlasztf} shows the long-term brightness behavior of \xmm, which varied between 18 - 21.5. Data points whose sky brightness is larger than the source brightness are omitted.

\subsection{\ztf observations}

\xmm was observed by the Zwicky Transient Facility (ZTF) \citep{masci+19} between May 2018 and July 2022. \textit{g} and \textit{i} filters are used in these observations. The brightness of the object was variable over a wide range between 17.3 and 21.6 mag. Hence, it was identified as the ZTF transient ZTF19aauymhu and classified as a likely CV  by the ALeRCE broker \citep{forster+21} but no follow-up work was initiated. The ZTF light curve is also shown in Fig.~\ref{f:atlasztf}.

\subsection{\pnstr observations}

\xmm was observed also by the Panoramic Survey Telescope and Rapid Response System (Pan-STARRS) \citep{chambers+16} as source id of PSO J152737.450-205305.091. Its brightness were obtained as \textit{m$_{\rm g}$}=20.25, \textit{m$_{\rm r}$}=18.86$\pm$0.1, \textit{m$_{\rm i}$}= 18.50$\pm$0.12, \textit{m$_{\rm z}$}=19.95$\pm$0.07 and \textit{m$_{\rm y}$}= 18.06$\pm$0.12 in different passbands.

\subsection{\glx observations}

The Galaxy Evolution Explorer (\textit{GALEX}) \citep{martin+05} was observed object as observation ID of 6383159947039868090. The ultraviolet brightness was measured as \textit{m$_{\rm NUV}$}= 21.99$\pm$0.46 in the near ultraviolet region (1750-2800 \AA) with 200 sec exposure time.

\subsection{\ero all sky survey observations}

The extended Roentgen Survey utilizing an Imaging Telescope Array (\ero) instrument \citep{predehl+21} on board the Spektrum-Roentgen-Gamma spacecraft \citep{sunyaev+21} detected \xmm in each sky survey. The data were downloaded from DATool\footnote{\protect\url{https://erosita.mpe.mpg.de/internal/DATool/}} which are a production of \ero Science Analysis Software System (\texttt{eSASS}) and the data version of the c947 was used for further analysis \citep{brunner+22}. The sky positions of the \xmm was measured in these scans as eRASS1 RA=231.9073(8) deg and DEC=-20.8856(8) deg, eRASS2 RA=231.906(1) deg and DEC=-20.885(1) deg, eRASS3 RA=231.906(1) deg and DEC=-20.885(1) deg and eRASS4 RA=231.9079(6) and DEC=20.8845(6), respectively. The uncertainties are given in the parenthesis with one sigma confidence for the last digits. The observation log of the relevant observations is displayed in Tab.~\ref{t:3}.

\begin{figure*}
\resizebox{\hsize}{!}{\includegraphics{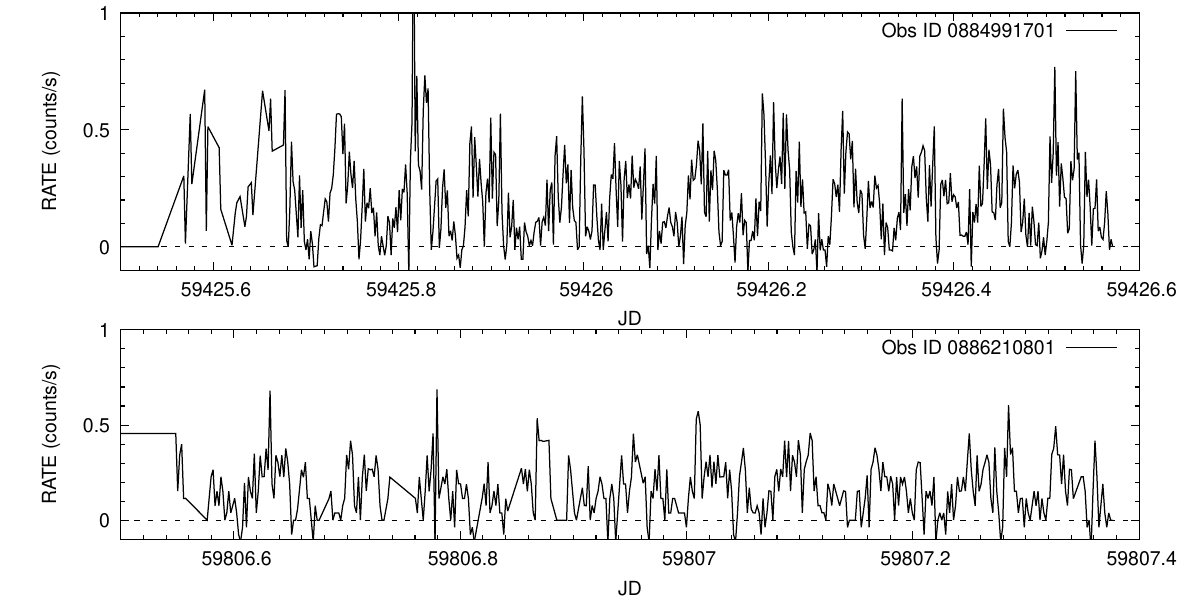}}
\caption{\xmmn light curve of \xmm obtained from EPIC-pn instrument. The light curve includes photons energies between 0.2 - 10 keV with 150 sec time bins.
\label{f:xmmlc}}
\end{figure*}

\begin{figure}
\resizebox{\hsize}{!}{\includegraphics{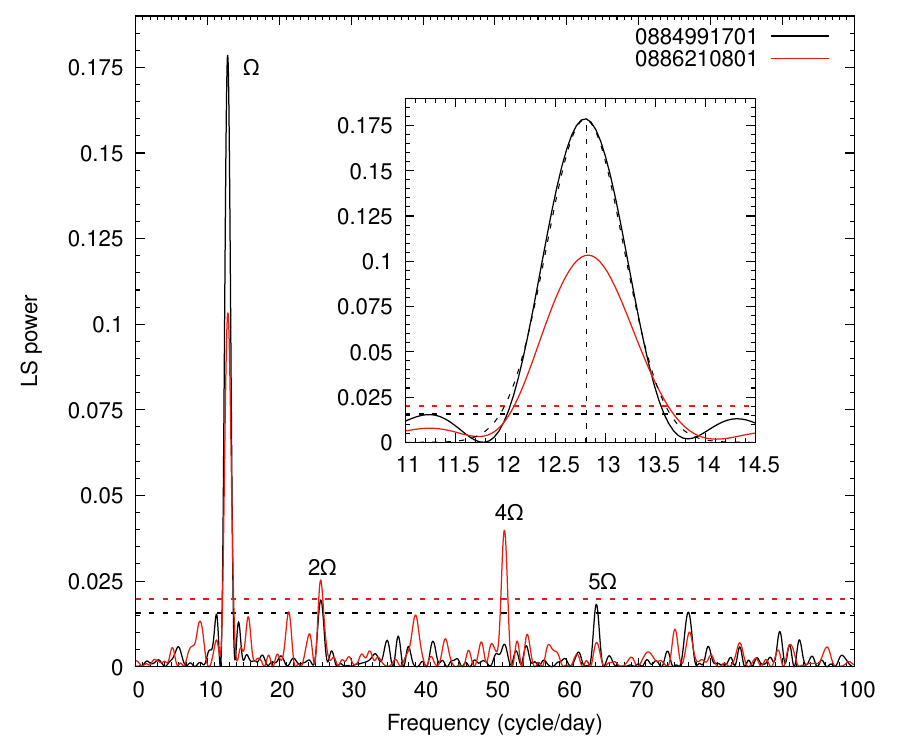}}
\caption{Power spectrum of the \xmm. The spectrum obtained from two different observations. Main figure shows the power spectra of the object which obtained from \xmmn at 0.2 - 10 keV with 50 sec timebins that are over plotted together with an inset detailing the main peak. Dashed black line shows a gaussian line fit to the main peak of the 0884991701 numbered observation. The black (Obs 1-0884991701) and red (Obs 2-0886210801) dashed lines in the main frame give the 95\% significance level of the both observation. The main period and its harmonics are indicated with the $\Omega$ symbol.}
\label{f:powspec}
\end{figure}

\subsection{SALT spectroscopic observations}

The optical spectroscopic identification of the source was carried out with the 11-meter Southern African Large Telescope (SALT) at Sutherland Observatory \citep{buckley+06}. Two 900-second exposures were obtained with the Robert Stobie Spectrograph \citep[RSS,][]{burgh+03} on 2023 April 26. The used grism PG0700 covers the wavelength range 3579 -- 7462\,\AA\, with two gaps between 4867 -- 4937\,\AA \; and 6195 -- 6255 \AA \; due to the chips gaps of the sensor. The used grism together with the used slit width of 1\farcs5 resulted in a spectral resolution of 7.8 \,\AA\,(FWHM). Standard \texttt{IRAF} \citep{tody1986iraf} procedures were followed in the reduction of the spectra which include dark- and bias-subtraction and flat-fielding. The spectrophotometric standard star LTT2415 \citep{oke90}, observed with SALT on 2023 May 1, was used for the flux calibration. The calibration was performed using the \texttt{STANDARD} task within \texttt{IRAF}, where the observation of the spectrophotometric standard star is compared to true flux data for the respective star within the \texttt{IRAF} database. This was followed by running the \texttt{sensfunc} task, which fits a function (spline3 in this case) to the measurement of the spectrophotometric standard star obtained in the previous task. Lastly, the task \texttt{CALIBRATE}, which uses the sensitivity function obtained in the previous step, was used to flux calibrate the spectra.

\section{Analysis and results \label{s:ana}}

\subsection{\xmmn observations}

\subsubsection{X-ray photometry \label{s:xobs}}

X-ray light curves of \xmm are shown in the original time sequence in Fig.~\ref{f:xmmlc}. The data were used in the interval $0.2 - 10$ keV and time bins of 50 seconds. The most striking feature of the X-ray light curve of the object is the succession of pulses or, in other words, consecutive hump-like structures. These humps individually exhibit a highly variable behavior. It is noticeable that each hump includes a deep and narrow dip-like minimum, and with wide minima between the humps. The wide minima can always clearly be recognized while the dip-like minimum is difficult to follow in individual humps due to the overall high degree of variability and relative faintness of the source.

We searched the X-ray lightcurves for possible strong periodic signals by computing Lomb \& Scargle periodograms \citep[][]{lomb+76,scargle+82}. The period analysis was performed using photons between 0.2 and 10 keV with 50 sec time bins. Significant peaks are searched in the frequency interval from 0 $d^{-1}$ up to the Nyquist frequency of the \xmmn data (480 $d^{-1}$). In both data sets, we obtained only one single peak at the same period of \textit{f}=12.81 cycles $d^{-1}$ (0.078 days) and its harmonics. The power spectra are shown in Fig.~\ref{f:powspec}.

\begin{figure}
\resizebox{\hsize}{!}{\includegraphics[width=\columnwidth]{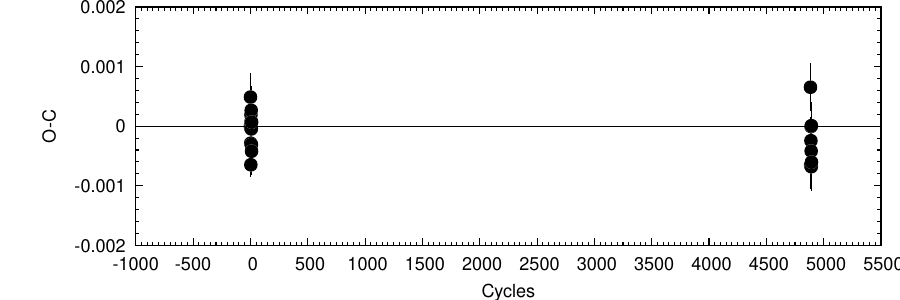}}
\caption{The plot of the residuals from minimum times of the deep dip after subtraction (O-C) of the linear trend. The solution for the corresponding linear fit is given in Eq.~\ref{e:eph}.
\label{f:oc}}
\end{figure}

\begin{figure}
\resizebox{\hsize}{!}{\includegraphics[width=\columnwidth]{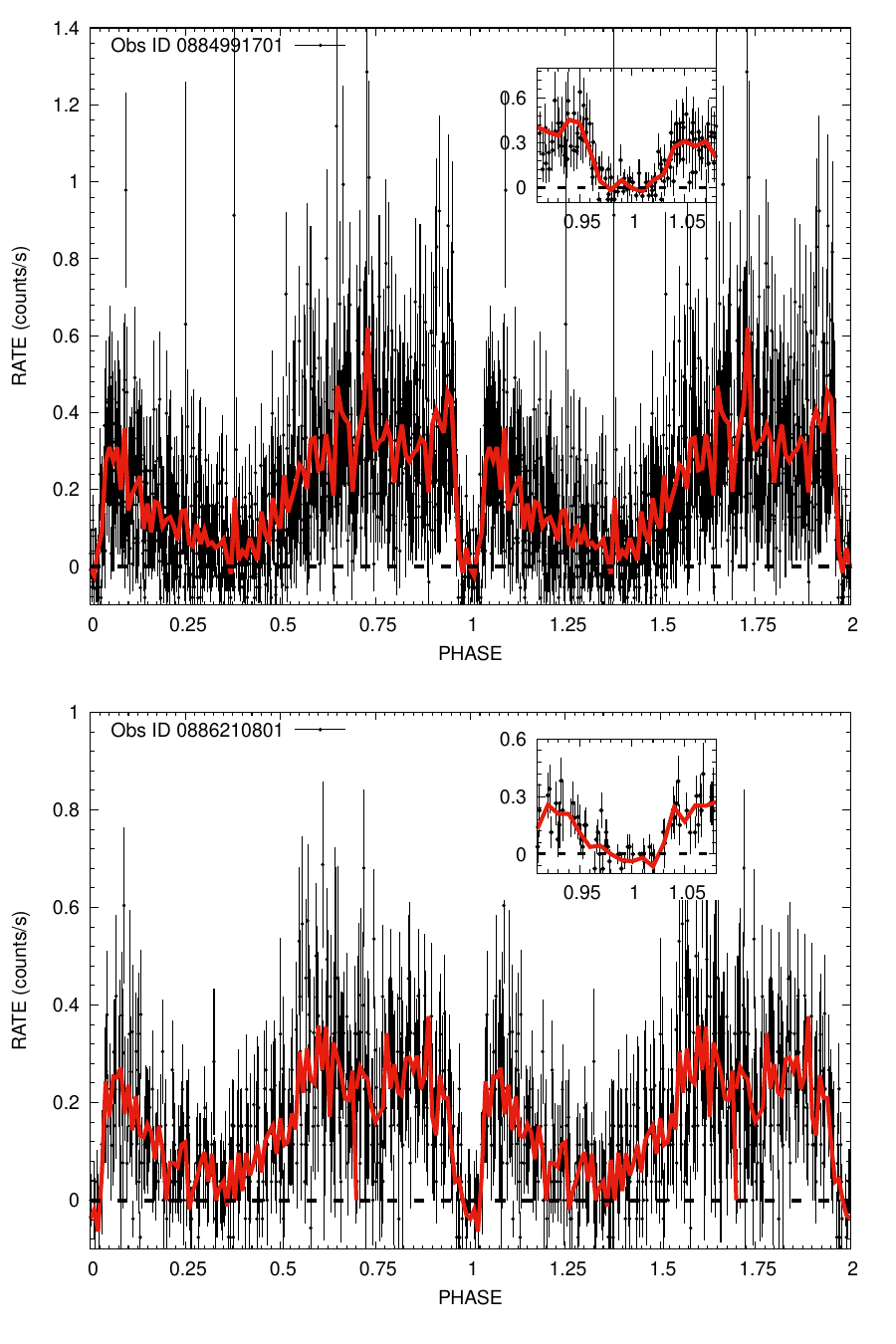}}
\caption{The folded X-ray light curves of the \xmm obtained in 0.2 - 10 keV energy range by \xmmn with time bins of 50\,s. The insets focus on deep dip interval. Bold red line shows the grouped phase bins with 0.01 phase units.
\label{f:foldlc}}
\end{figure}

The deep dip, particularly noticeable in the complex structure of the light curve, exhibits apparent boundaries and demonstrates a sharp decline. The obtained timings from these distinct edges were used to ascertain the precise center of the dip and enhance the period acquired from the periodogram. For this process, the ingress ($T_{\rm i}$) and egress ($T_{\rm e}$) times were determined along the light curve from the light curves obtained in the 30-second time bin in the deep dip. The selection of phase zero was established at the midpoint ($T_{\rm min}$) of these $T_{\rm i}$ and $T_{\rm e}$ times from both \xmmn observations. The regression was first performed for the 12 continuous photometric cycles obtained from Obs 1. The period obtained after regression of the first observation was measured as 0.07799 $\pm$ 0.00003 days. The initial regression has an inaccuracy of 2.9 seconds. If the regression error is obtained by Obs 1 interpolated to Obs 2, i.e. 380.77 days later, the accumulated error can produce a cycle count error with 2.1 cycles. Although this suggests that both observations should be evaluated separately, we also consider it appropriate to include the regression including Obs 1 and Obs 2 together here, as this error may be caused by the low signal-to-noise ratio data. These measurements are shown in Tab.~\ref{t:min}. Due to the uncertainty in the number of cycles for the second observation, the number of cycles are indicated with a colon symbol for Obs 2. Thus, a weighted linear regression between the times $T_{\rm min}$ and the cycle numbers were used to derive

\begin{equation}
BJD(T_{0,min}) = 2459425.6822(8) + 0.07798799(5) \times E
\label{e:eph}
,\end{equation}
where the numbers in parentheses give the uncertainties in the last digits. In addition to this regression, we find it appropriate the highlight the periods that may cause $\pm$1 and $\pm$2 cycles ambiguities in the final period. The periods that can be obtained as a result of $\pm$1 cycle difference are 0.077972 days and 0.078002 days, while the periods that can be obtained as a result of $\pm$2 cycle difference are 0.077955 days and 0.078020 days, respectively. Zero time refers to the center of the dip or eclipse. The residuals of this linear fit are shown in Fig.~\ref{f:oc}. In the following, all phases refer to the ephemeris given in Eq.~\ref{e:eph}. We used this ephemeris to produce phase-folded light curves. The figure presented in Fig.~\ref{f:foldlc} displays folded light curves obtained from the \xmmn observatory. The narrow dip and hump-like structures are similarly observed in both light curves. Both datasets indicate a single period based on our timing analysis. Such behavior is consistent with the $P_{\rm orbit} = P_{\rm spin}$ behavior seen in polar-type magnetic CVs. The identified period is likely the orbital period.

\begin{figure}
\resizebox{\hsize}{!}{\includegraphics[width=\columnwidth]{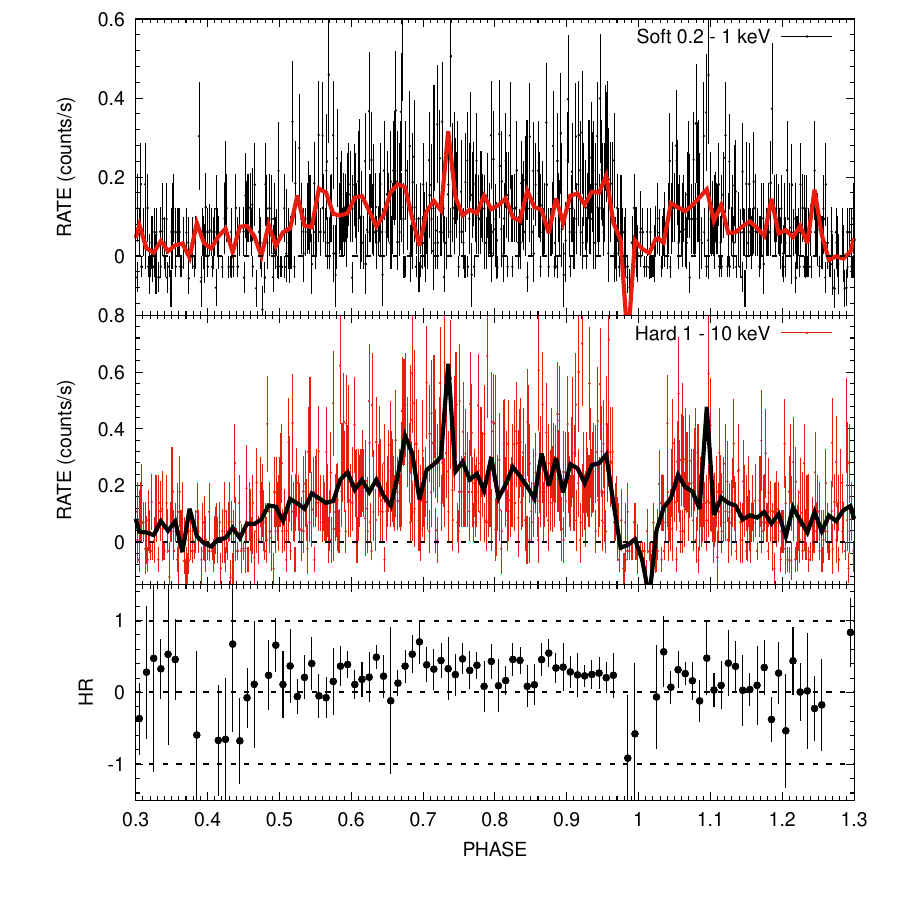}}
      \caption{Energy resolved light curves and HR variation of the \xmm with the Obs ID of 0884991701 (Obs 1). These phase folded light curves are produced from 100 s time binnned time series. 
\label{f:foldlceng1}
} 
\end{figure}

\begin{figure}
\resizebox{\hsize}{!}{\includegraphics[width=\columnwidth]{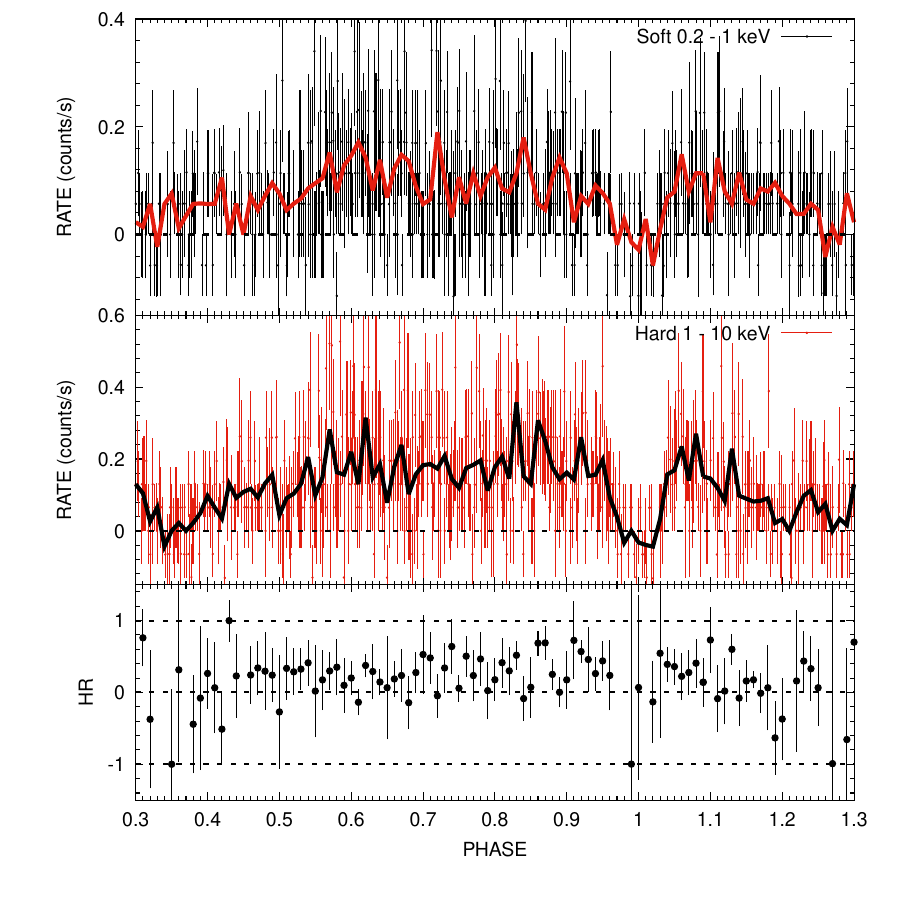}}
      \caption{Energy resolved light curves and HR variation of \xmm with the Obs ID of 0886210801 (Obs 2). These phase folded light curves are produced from 100 s time binnned time series. 
\label{f:foldlceng2}
} 
\end{figure}

We generated energy-resolved and phase-folded X-ray light curves in two energy bands, a soft band between 0.2 - 1 keV and a hard band between 1 - 10 keV. These are shown in Fig.~\ref{f:foldlceng1} and Fig.~\ref{f:foldlceng2}. Within the errors, both dips have a similar width. The dip or eclipse has a length of 9.2$\pm$1.2 min and 9.1$\pm1.4$ min for Obs 1 and Obs 2, respectively. The errors in the eclipse lengths are derived from the standard deviation of these individual measurements. When considering errors, both observations indicate the net count rates are compatible with zero in the dips. The second wide minimum, on the other hand, always has positive counts. The mean count rates obtained from the 0.2 - 0.4 phase interval are 0.014$\pm$0.002 count s$^{-1}$ and 0.05$\pm$0.01 count s$^{-1}$ for Obs 1 and Obs 2, respectively. 

The center of the vast minimum is located at phase 0.35 (see Fig.~\ref{f:foldlc}). The mean count values obtained from 0.65 - 0.9 phase interval (bright phase) in Obs 1 is 0.30$\pm$0.01 counts s$^{-1}$  and 0.24$\pm$0.01 counts s$^{-1}$ for Obs 2. The difference is 1.25 factor but morphologically, both light curves are identically the same.

We calculated the Hardness Ratio (\textit{HR}) between the two bands to understand better deep dip-like features and possible small variations which are defined as $HR = (H-S)/(H+S)$ and hence vary between $-1$ and $+1$ (see Fig.~\ref{f:foldlceng1}). In Obs 1, between two dips, the HR varies slightly and spreads around 0.2 during bright hump. Negative counts in the center positions of the dips prevented the calculation of meaningful HR values. At the same time, no significant hardening was detected in the ingress and egress of the dips which could be related to mass transition or absorption. In Obs 2, HR is a little more variable. However, as in the first observation, it scatters around 0.2, and similarly, there is no sign of hardening related to an element that can create absorption in the ingress and egress parts of the dip (see Fig.~\ref{f:foldlceng2}). Given that both observations include negative counts, it is not practical to do a significant calculation in vast and deep minima. Therefore, the HR values in these intervals were outside the accepted range of variation. However, it is important to note that no evidence of a hardening or absorption condition, which may suggest the presence of mass flow at the ingress and egress of the deep minimum, has been detected.

\subsubsection{X-ray spectroscopy \label{s:hr}}

Spectral analysis was performed with the XSPEC package \citep[version 12.12.0][]{arnaud+96,dorman+01}. Only photons between 0.2 and 10 keV were included in the spectral analysis. We started the spectral analysis by considering only the bright phase in the light curve between $\phi = 0.65 - 0.90$ phases (see Fig~\ref{f:foldlc}). We assume that the bright phase will contain the physical status of the main radiation region on the source. Photons were grouped with a minimum of 20 counts per bin and we used $\chi^2$ statistics for optimizing the fit.

\begin{figure}
   \includegraphics[width=\columnwidth]{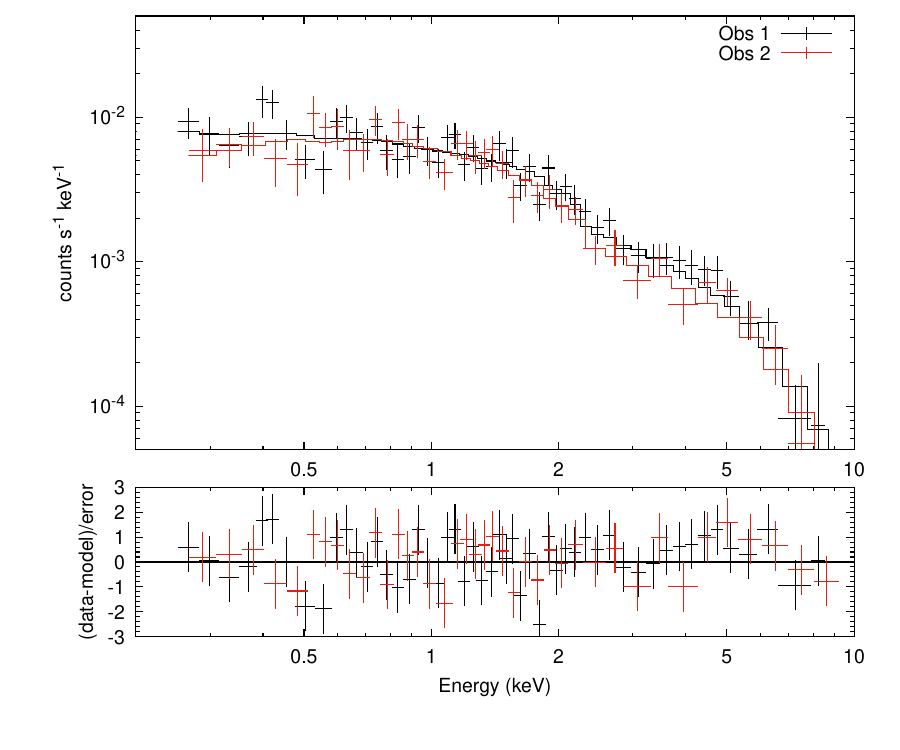}
      \caption{Mean bright-phase spectra of \xmm for Obs 1 and Obs 2. The bright phase spectra obtained between 0.65 and 0.90 photometric phases according to Eq.~\ref{e:eph}. }
      \label{f:brightpha}
\end{figure}

Since the count values of the object were higher in Obs 1, we started the spectral analysis with this data. Initially we used a single collisionally ionized {\tt APEC} emission model \citep{smith+01} with solar abundances \citep{wilms+00} absorbed by some cold interstellar matter with hydrogen column density ($N_{\rm H}$) so that our model reads {\tt TBABS * APEC} in {\tt XSPEC}. This basic model qualitatively describes the spectrum fairly well. The model gave the $\chi^2 = 52$ for 50 degree of freedom $\chi^2_\nu = 1.04$. Interestingly, the temperature value is pegged at the maximum for the {\tt APEC} model.  We used \textit{steppar} command in {\tt XSPEC} to see any possible statistical minima between 0.05 and 64 keV. The statistical minimum was in a decreasing trend, implying temperature is higher than 64 keV and it was not possible to detect the exact minimum point. The difficulty in detecting the spectrum temperature is thought to be connected to the photon contribution in the hard X-ray region ($>5$~keV) being more than predicted. Particularly in terms of magnetic CVs, pegging at the maximum allowed temperature is typical. The real temperature should be much lower but the spectrum seems to be that hot (harder) due to the reflection, one should be considered the main emission model with the reflect model or partial covering fraction absorption ({\tt PCFABS}) \citep[see for more information,][]{mukai17,schwope+20}. Therefore, we replaced the absorption component with a partial covering fraction absorption ({\tt PCFABS}) and reapplied the fit. This simple model gave a very good fit and {\tt XSPEC} measured a lower temperature as $kT_{\rm apec}$= 21 keV (see Fig.~\ref{f:brightpha}). The final fit is yielded with the $\chi^2 = 52$ for 49 degrees of freedom $\chi^2_\nu = 1.06$.

\begin{table}
\centering
\caption{Minimum times, cycles and $\Delta$T times of the \xmm. The cumulative error obtained from regression from the Obs 1 indicates that the second observation might result in a variance of $\pm$2 cycles in the cycle count. Thus, the cycles of the second observation are denoted by colon.}
\label{t:min}
\resizebox{6cm}{!}{
\begin{tabular}{ccc}
Minimum times(BJD) & Cycles & $\Delta$T\\
\hline
\hline 
2459425.6822(4) & 0 & 7.28$\times10^{-12}$ \\ 
2459425.7597(4) & 1 & 4.88$\times10^{-4}$ \\ 
2459425.8382(2) & 2 & $-$2.40$\times10^{-5}$ \\ 
2459425.9165(4) & 3 & $-$2.86$\times10^{-4}$ \\ 
2459425.9948(2) & 4 & $-$6.48$\times10^{-4}$ \\ 
2459426.0720(2) & 5 & 1.90$\times10^{-4}$ \\ 
2459426.1501(4) & 6 & 7.80$\times10^{-5}$ \\ 
2459426.2279(4) & 7 & 2.66$\times10^{-4}$ \\ 
2459426.3062(4) & 8 & $-$4.60$\times10^{-5}$ \\ 
2459426.3844(4) & 9 & $-$3.08$\times10^{-4}$ \\ 
2459426.4625(4) & 10 & $-$4.20$\times10^{-4}$ \\ 
2459426.5400(2) & 11 & 6.79$\times10^{-5}$ \\ 
2459806.6529(4) & 4885: & 6.54$\times10^{-4}$ \\ 
2459806.7322(4) & 4886: & $-$6.58$\times10^{-4}$ \\ 
2459806.9658(4) & 4889: & $-$2.44$\times10^{-4}$ \\ 
2459807.0435(4) & 4890: & $-$5.92$\times10^{-6}$ \\ 
2459807.1219(4) & 4891: & $-$4.18$\times10^{-4}$ \\ 
2459807.2002(4) & 4892: & $-$6.80$\times10^{-4}$ \\ 
2459807.2775(4) & 4893: & 8.07$\times10^{-6}$ \\ 
2459807.3561(4) & 4894: & $-$6.04$\times10^{-4}$ \\ 
\hline 
\hline 
\end{tabular}
}
\end{table} 

We tried to model the spectrum for the bright phase by using the same spectral model for Obs 2. We could not obtain any meaningful temperature values in any of our trials with {\tt PCFABS}. Similarly in the previous observation, the temperature value was entrapped at maximum value. In the parameter test we conducted with \textit{steppar} command, we found that the temperature entered the 90\% confidence value limit after 11 keV. However, we could not see any dip point of the statistical minimum in the temperature range defined for {\tt APEC}. Then we tried to change our absorption method and changed {\tt PCFABS} to the {\tt TBABS}. The fit was re-performed again and finally, we were able to calculate a temperature, albeit with large error margins. The result obtained with this model is the $\chi^2 = 40$ for 38 degrees of freedom $\chi^2_\nu = 1.05$. The best-fit parameters for the bright phase spectra are shown in Tab.~\ref{t:2}.

One of the most interesting results from Obs 1 is the high hydrogen column density. The bright phase spectra obtained in the high state are significantly higher than the galactic column density. In the direction of the object, the galactic column density is $N_{\rm H,gal}$=$8.32\times10^{20}$\,cm$^{-2}$ according to \cite{bekhti+16}. According to this result, the column density for Obs 1 appears to be a factor of 18 higher than the galactic column density. This is quite high. But we want to emphasize that even the lower error limit of this parameter, 0.45 (factor of $\approx$6), is considerably higher than the galactic column density. Instead, a difference of 0.3 factor is seen between the column density derived from Obs 2. The analysis suggests that the spectrum acquired during the high state and bright phase, particularly when the accretion region is at its maximum visibility could be affected by certain absorption features. Given the current accretion condition of the object, it is plausible that it could be attributed to either a stellar wind or an accretion curtain, or perhaps an extended part of an accretion stream. The model combining absorption components ({\tt TBABS * PCFABS * APEC}) was also evaluated for Obs 2. However, no significant results were obtained in measuring the parameters of the partial absorption component. The total column density calculated using {\tt TBABS} is comparable to the model that does not include a partial absorption.

\bgroup
\def\arraystretch{1.10}
\begin{table}
\centering
\caption{Spectral parameters from bright phase interval ($\phi$=0.65 - 0.90): spectral fit parameters, their uncertainties, fit statistics, and model bolometric fluxes.}
\label{t:2}
\resizebox{\columnwidth}{!}{
\begin{tabular}{lcc}
\hline       
\hline                     
\multicolumn{2}{l} {Obs 1 - Model: {\tt PCFABS*(APEC)}}    \\
\hline
Parameters               &                                 \\
\hline
   
$N_{\rm H}$($10^{22}$cm$^{2}$)  & $1.41^{+1.35}_{-0.79}$   \\
CvrFract                        & $0.45^{+0.12}_{-0.13}$   \\
$kT_{\rm apec}$(keV)            & $21^{+19}_{-11}$         \\

$\chi^{2}$(d.o.f)            & 1.06(52/49)                 \\
\hline
Unabsorbed Fluxes $(10^{-13}\fergs)$                       \\
\hline
$F_{0.5-2.5} $  & $0.91^{+0.79}_{-0.09}$        \\
$F_{2.5-10}  $  & $2.35^{+0.31}_{-0.39}$        \\
$F_{\rm bol}     $  & $5.8^{+1.5}_{-1.5}$       \\
\hline
$L_{x}$(erg s$^{-1}$)$(10^{31})$ & $4.6\pm1.2$       \\
\hline
\hline                     
\multicolumn{2}{l} {Obs 2 - Model: {\tt TBABS*(APEC)}}      \\
\hline
Parameters               &                                  \\
\hline
   
$N_{\rm H}$($10^{22}$cm$^{2}$)  & $0.03^{+0.03}_{-0.02}$    \\
$kT_{\rm apec}$(keV)            & $46^{+14}_{-11}$          \\

$\chi^{2}$(d.o.f)            & 1.05(40/38)                  \\
\hline
Unabsorbed Fluxes $(10^{-13}\fergs)$                        \\
\hline
$F_{0.5-2.5} $  & $0.6^{+0.7}_{-0.6}$         \\
$F_{2.5-10}  $  & $1.54^{+0.13}_{-0.23}$      \\
$F_{\rm bol}     $  & $6.3^{+0.85}_{-0.50}$   \\
\hline
$L_{x}$(erg s$^{-1}$)$(10^{31})$ & $5.0\pm0.8$  \\
\hline
\end{tabular}
}
\end{table} 
\egroup

The spectra seen in the two observations exhibit a notable absence of the blackbody extension in the soft X-ray regime. Many polars have a soft X-ray radiation component in the X-ray spectrum. The photosphere of the white dwarf located under the shock can be exposed to radiation from above and heated by shock region, resulting in the formation of a highly concentrated soft X-ray source \citep{mukai17}. While the absence of the spectral extension associated with this emission is not indicative of the object being non-polar, it typically appears in polars. The detection of this emission was not evident in serendipitously discovered polars by \xmmn observations. The narrow coverage of the energy area of the instrument or suppression of the accretion region by an absorber is regarded to be the underlying cause for its non-existence.


The unabsorbed fluxes in different bands were calculated using the best-fit models for both observations. The \textit{cflux} command in XSPEC was used for this. For a more comprehensive total X-ray flux calculation in models, we create dummy responses using the \xmmn response matrices loaded into {\tt XSPEC} and the unabsorbed bolometric flux was calculated in the range 10$^{-6}$ - $10^3$\,keV with reference to the corresponding model parameters. The correction factors ($F_{\rm bol}$/$F_{\rm 0.5-10}$) for the total X-ray flux in both luminosities are 1.77 and 2.94, respectively. The results of these calculations and their uncertainties are shown in Table~\ref{t:2}.

The determination of the mass accretion rate of the object involves the computation of the total X-ray flux derived from an ideal model fitted throughout the X-ray spectrum. This calculation assumes a correlation between the observed emission and the amount of energy produced during the accretion process ($L_{\rm acc} \approx L_{\rm X}$). In further elaboration, it is essential to include the non-thermal cyclotron emission ($F_{\rm cyc}$) generated by the motion of matter within a magnetic field, in addition to X-rays ($F_{\rm acc} \approx $ $F_{\rm X}+F_{\rm cyc}$). However, the strength of this radiation strongly depends on the geometry and can vary significantly depending on the orbital inclination and the spatial arrangement of the accretion structures which cannot be precisely identified with the available data. Hence, the contribution of this emission is omitted in the present computation.

The distance of the object is given as 1156 pc by the Gaia DR3 catalog. Using this distance, with a white dwarf mass of 0.8 $M_{\odot}$, assumed to be the average for the CVs \citep{littlefair+08, savoury+11, pala+22}, and a radius calculated from \cite{nauenberg+72} concerning this mass, we calculate the X-ray luminosity for Obs 1 as $L_{\rm X}$=4.6$\pm 1.2 \times10^{31}$ \lergs. Thus, when the X-ray parameters obtained from the bright phase spectrum and the distance of the object are taken into account, the mass accretion rate in Obs 1 is $\dot{M}=5.0\pm 1.0$ $\times10^{-12}$ $M_{\odot}$ yr$^{-1}$. When we perform similar calculations for the data obtained from Obs 2, the X-ray luminosity is $L_{\rm X}$=5.0$\pm0.8 \times10^{31}$ \lergs and the mass accretion rate is $\dot{M}=5.3 \pm 0.7$ $\times10^{-12}$ $M_{\odot}$ yr$^{-1}$. From these calculations, it can be assumed that for a fixed distance of 1156 pc, the mass accretion between the two \xmmn observation, the accretion rate of \xmm varied between (4 - 6) $\times 10^{-12} M_{\odot}$ yr$^{-1}$ within the errors. We only took into account the errors in fluxes in these calculations. Considering the distance error derived from \gai DR3 i.e. 817 pc and 1876 pc as upper and lower limits, it can be stated that $\dot{M}$ lies between the range of (3 - 13) $\times10^{-12}$ $M_{\odot}$ yr$^{-1}$. Both observations indicate that \xmm has similar accretion rates.


\subsection{Long-term optical photometry}

The long-term optical photometry of \xmm is shown in Fig.~\ref{f:atlasztf}. While ATLAS has the longest duration of observation, it also encompasses the temporal range of observation presented by the ZTF survey. The observed data demonstrate comparable patterns over the same time interval of convergence. Within the ATLAS survey, the brightness of the object exhibits a variation stretching from 17.9 to 21.4 mag in \textit{c} band (i.e. 3.5 magnitude difference). The ZTF, which operates with Sloan Digital Sky Survey (SDSS) filters, has recorded brightness variations in the range 18-20.6 mag in z$_{r}$ band. Furthermore, Fig.~\ref{f:atlasztf} illustrates the match in the first (Obs 1) \xmmn observation and the corresponding optical data.


We found no X-ray behavior pattern in the optical light curves when compared to the photometric period. This phenomenon may have several causes. The large optical variations in the ATLAS and ZTF light curves may be due to orbital motion or various mass accretion states. Inadequate light curve sampling or period inaccuracy from the first observation may have caused a cycle alias in Obs 2, similar to the optical observations, leading to misplaced photometric data points. One or more of these may in progress for this incompatibility. Hence, we cannot comment on the phase-dependent orbital variability in the optical wavelength range. No dedicated time-resolved optical photometry could be performed yet and the accuracy of our ephemeris is not sufficient to construct a phase-binned light curve from the sparsely sampled ZTF or ATLAS data. 


\begin{figure}
   \includegraphics[width=\columnwidth]{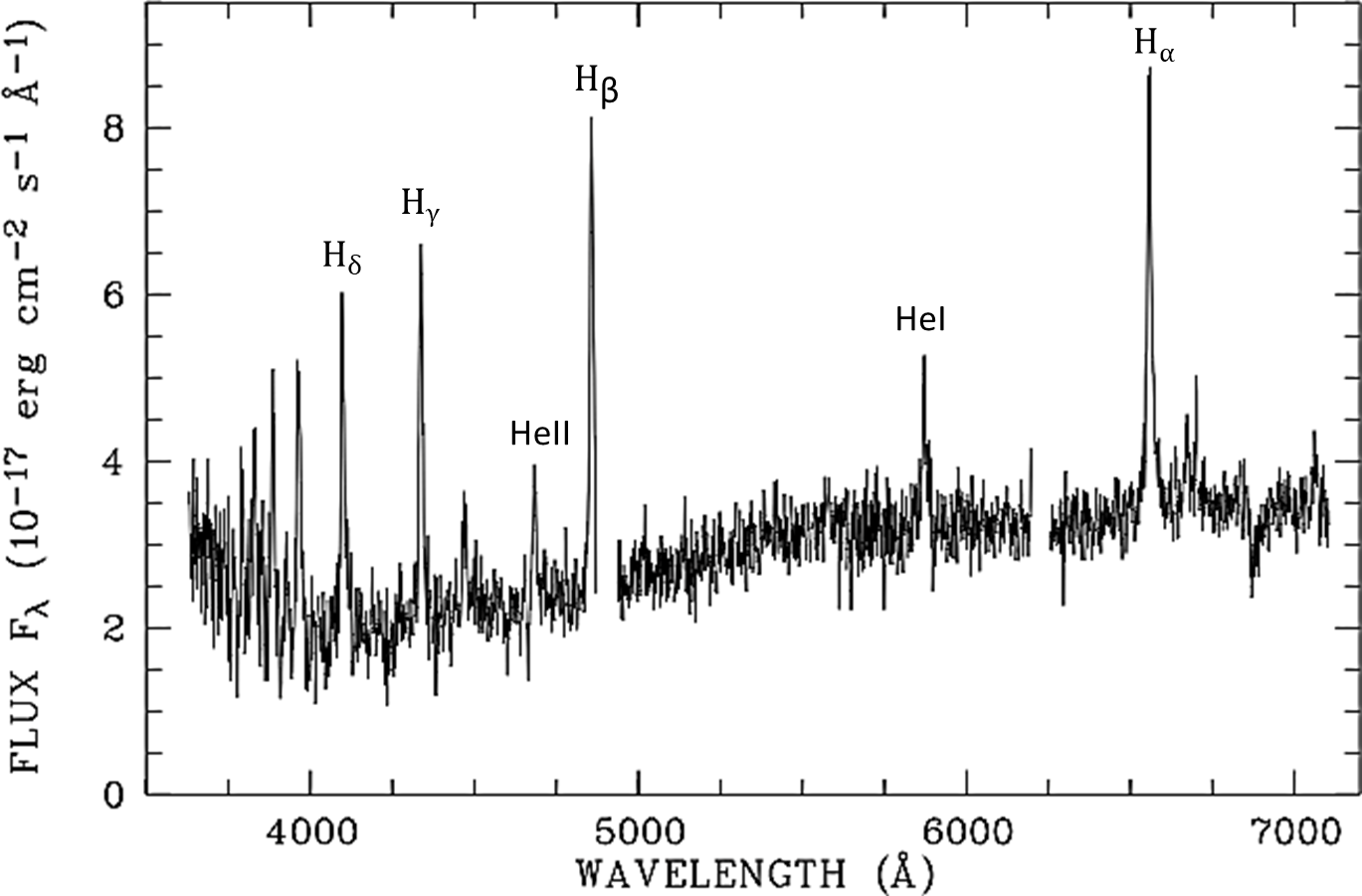}
      \caption{The SALT spectrum of \xmm. } 
      \label{f:salt}
\end{figure}

\subsection{SALT spectroscopy}

Figure \ref{f:salt} shows the average combined flux calibrated spectrum of the two SALT observations obtained of XMM J152736. The gaps seen in the spectrum between 4867 -- 4937\,\AA \;and 6195 -- 6255\,\AA \; are from the chip gaps in the sensor. The spectrum shows very prominent Balmer emission lines, although H$\beta$ (at 4861\,\AA)\; is partially cut off due to the occurrence (or overlap) of the chip gap. Apart from the Balmer lines, emission lines of He I, at  $\lambda$$\lambda$ 4472, 5876 and 6678\,\AA, as well as He II $\lambda$4686\,\AA. 


\begin{figure}[h]
\includegraphics[width=\columnwidth]{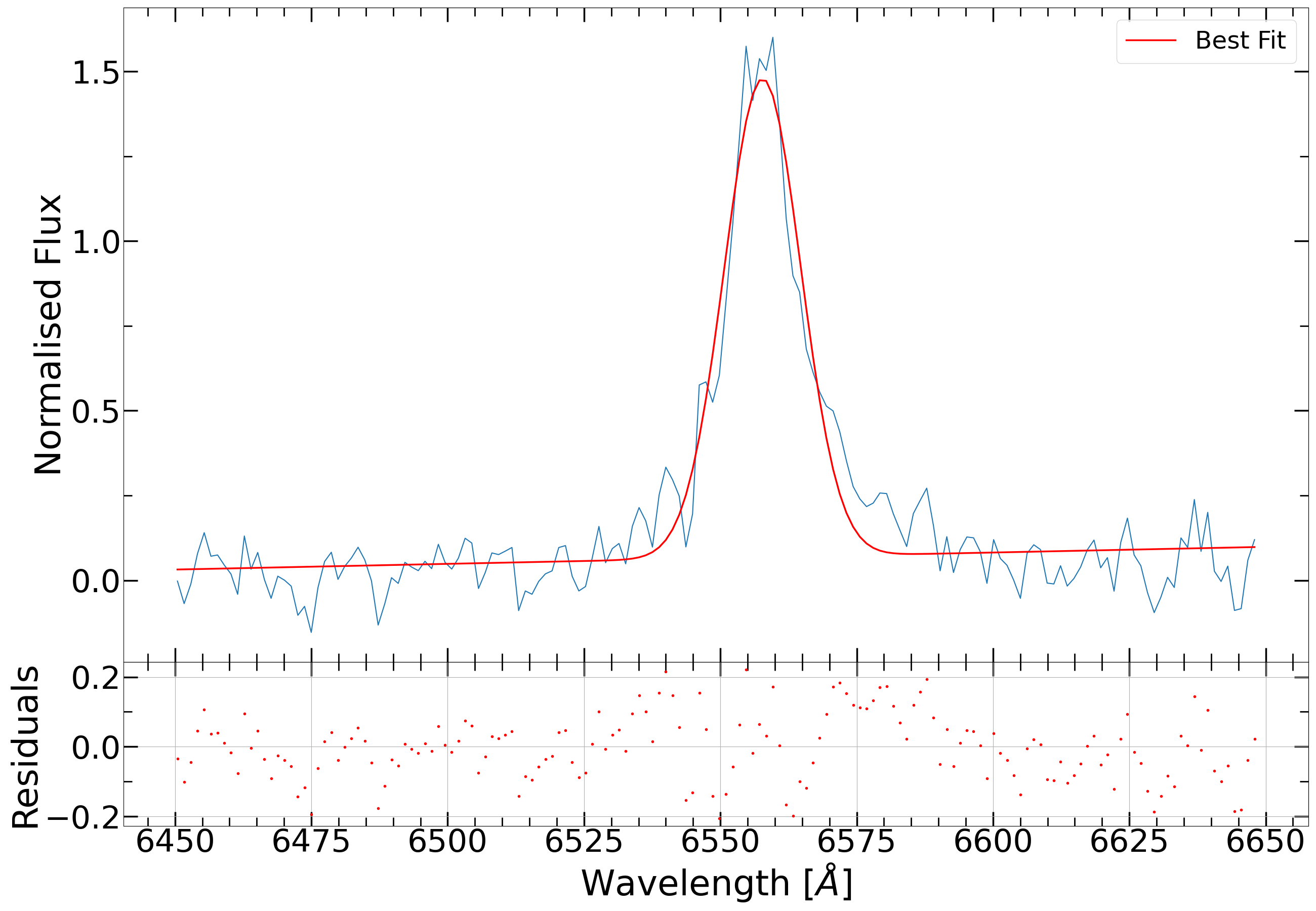}
\caption[H$\alpha$ \texttt{lmfit} modelling]{The \texttt{lmfit} modelling to the spectral region around the H$\alpha$ emission line. The fit consists of a straight line approximating the continuum and a Gaussian for the line itself. Shown is the two-component fit with a red line and the data in blue in the upper panel, while the bottom panel shows the residuals of the fit.
}
\label{f:hafit}
\end{figure}

Close inspection of the emission line profiles, especially the Balmer series, also shows possible indications of multiple components, however, this might also be due to noise in the data. Measurements of the most prominent emission lines were obtained using the Python package \texttt{lmfit} \citep{2016ascl.soft06014N} on the average combined spectrum from the two SALT observations. Before measuring the spectral lines, however, this average combined spectrum was first continuum normalized using the \texttt{numpy} Python package by dividing the spectrum by a fourth-degree polynomial that was fitted to the continuum. Figure \ref{f:hafit} shows an example of the \texttt{lmfit} modeling to H$\alpha$, while Table \ref{t:linpar} shows the radial velocity as well as the FWHM of the most prominent emission lines. 

He I $\lambda$6680 could not be accurately modeled using \texttt{lmfit} due to high noise levels in the spectrum. All of the lines show blue shifted radial velocities, indicating that high-resolution time-resolved spectroscopic observations could be viable to possibly obtain additional parameters such as the orbital period of the system. The equivalent width (EW) was measured by fitting a Gaussian profile to the respective line. The uncertainty of EW was derived by determining the uncertainty in the flux value, which was derived from the uncertainties in the full width at half maximum (FWHM) and the peak flux. These were found from the lmfit measurements to the line profile. A possible uncertainty in the continuum was neglected.


The continuum is rising toward the red wavelength range with maxima around 6600\,\AA\, and 5700\,\AA. This is reminiscent of a cyclotron continuum originating from a moderate magnetic field strength, perhaps similar to that seen in V834 Cen \citep{schwope+beuermann90}. Whether the two mentioned maxima represent individual cyclotron harmonics is difficult to decide given the rather modest wavelength coverage of the spectra and the lack of time resolution, i.e., phase coverage. However, both the presence of high excitation emission lines and the shape of the continuum strengthen the polar interpretation of the system.

\begin{table*}[h]
\centering
\caption{Optical emission line properties of the \xmm.}
\label{t:linpar}
\begin{threeparttable}
\begin{tabular}{cccc}
\hline
\hline
Line    & Radial Velocity &   FWHM    &   EW  \\
        &   (km s$^{-1}$)   &  (km s$^{-1}$)     &   (\AA)  \\
\hline
H$\alpha$   &   -236 $\pm$ 24   &   434 $\pm$ 58    &   32.2 $\pm$ 1.7 \\
H$\gamma$   &   -257 $\pm$ 23   &   540 $\pm$ 55    &   30.6 $\pm$ 1.6 \\
H$\delta$   &   -231 $\pm$ 30   &   666 $\pm$ 72    &   38.8 $\pm$ 1.9 \\
H$\epsilon$ &   -510 $\pm$ 34   &   745 $\pm$ 80    &   42.5 $\pm$ 2.2 \\
He I $\lambda$5876  &   -181 $\pm$ 12   &   874 $\pm$ 27    &   12.3 $\pm$ 0.7 \\
He I $\lambda$4472  &   -179 $\pm$ 34   &   687 $\pm$ 79    &   10.6 $\pm$ 1.8 \\
He II $\lambda$4686 &   -135 $\pm$ 11   &   820 $\pm$ 27    &   12.9 $\pm$ 0.6 \\
\hline
\hline
\end{tabular}%
\end{threeparttable}
\end{table*}


\subsection{eROSITA All Sky Survey}

\begin{figure}
   \centering
   \includegraphics[width=\columnwidth]{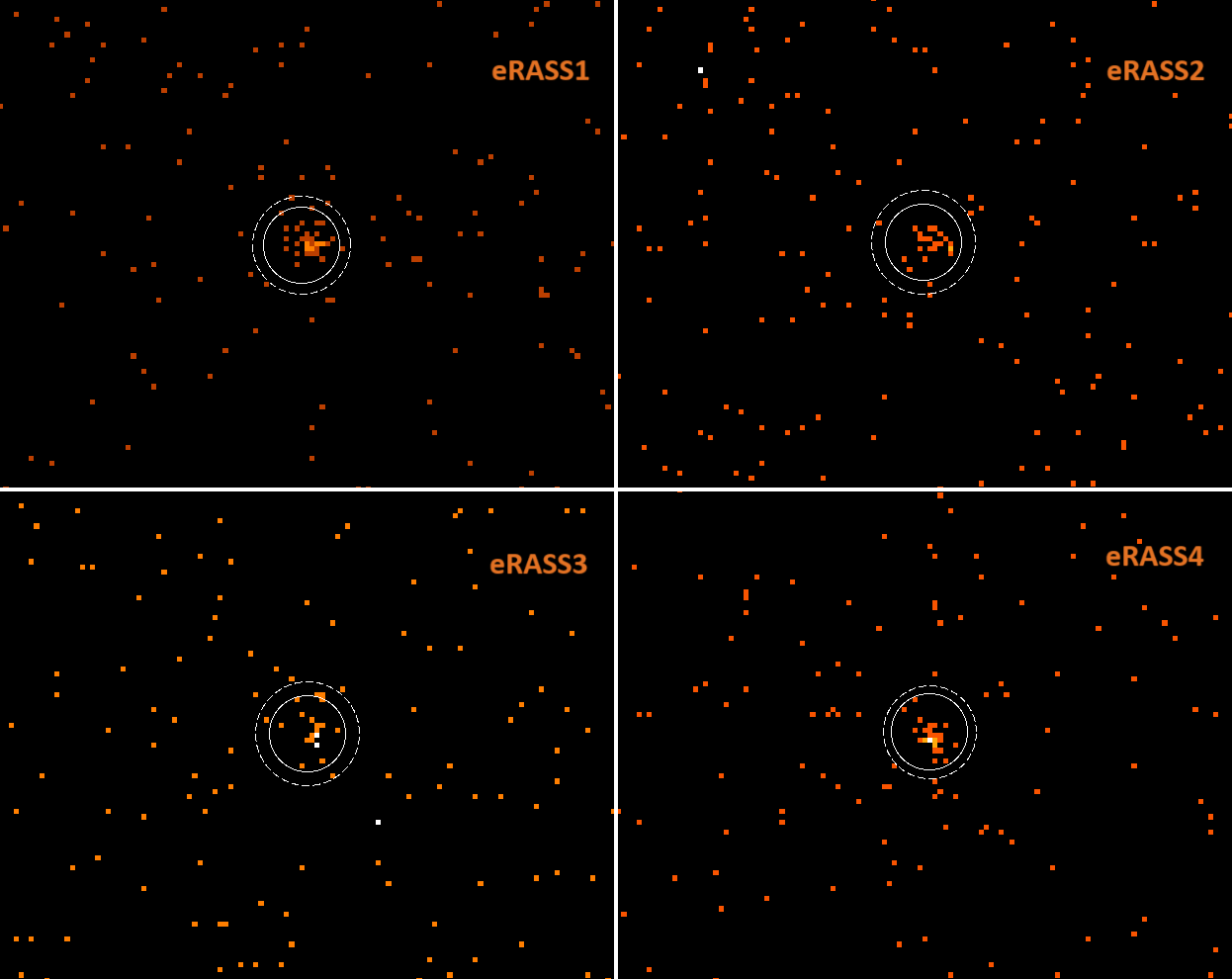}
      \caption{Images of the \xmm during the 4 \ero All Sky Surveys. The images obtained in the 0.2 – 10 keV energy range. The white apertures are positioned according to the eRASS1-4 detections with 30 arcsec. The dashed white circles show the uncertainties of these detections within 3$\sigma$ confidence.} 
      \label{f:erass14}
\end{figure}

The survey images obtained from eROSITA are shown in Fig.~\ref{f:erass14}. \xmm was detected in all surveys. In eRASS1(0.36$\pm$0.06 count/s) and eRASS4 (0.26$\pm$0.05 count/s) the object has relatively higher counts than others, in eRASS2 (0.15$\pm$0.03 count/s) and eRASS3 (0.16$\pm$0.04 count/s), low counts or lower energies were detected. As can be seen from Fig.~\ref{f:atlasztf}, there is no overlap with the long-term photometric data. In addition, we would like to point out that eRASS4 and \xmm-Obs 1, which are the closest points to each other, maybe in a similar accretion state considering the photometric data points.

As in \xmmn, we obtained the HR values of the system for all sky survey as HR1= 0.33$\pm$0.23 (eRASS1), HR2= 0.18$\pm$0.15 (eRASS2), HR3= 0.34$\pm$0.15 (eRASS3) and HR4= 0.05$\pm$0.19 (eRASS4) for the energy ranges 0.2 - 1.0 keV and 1.0 - 10 keV, respectively. Given the errors, it can be inferred that the HR, which refers to the spectral shape, behaves similarly to the bright phase in \xmmn observations.

\begin{table}
\centering
\caption{Observation log of the \ero All Sky Surveys and pointed observations covering the location of \xmm. The given times (MJD) refer to the middle of the concerned observation. The net count rates were obtained in 0.2 - 10 keV energy range. }
\label{t:3}
\resizebox{\columnwidth}{!}{
\begin{tabular}{lllc}
Survey &  Count Rates (s$^{-1}$)   & Time (MJD) & Total Exposure (s)\\
\hline
\hline 
eRASS1       & 0.355$\pm0.061$& 58902.836 & 111     \\
eRASS2       & 0.150$\pm0.036$& 59082.290 & 143     \\
eRASS3       & 0.160$\pm0.042$& 59251.562 & 119     \\
eRASS4       & 0.261$\pm0.050$& 59441.375 & 120      \\
XMM-Newton (Obs 1)   & 0.0173$\pm0.0006$& 59426.071 & 110ksec    \\
XMM-Newton (Obs 2)   & 0.0143$\pm0.0007$& 59806.842 & 106ksec    \\

\hline 
\hline 
\end{tabular}
}
\end{table} 

\subsection{Status of the Eclipse-like Feature and Accretion Geometry}

The prominent dip-like structure observed in the X-ray light curve has two possible explanations. Either it is caused by an eclipse of the white dwarf by the donor star or an obscuration of the accretion region on the white dwarf by the accretion stream (stream dip). When discussing its likely nature we are restricted to X-ray data only, time-resolved optical photometry would facilitate an interpretation but is not available. We first re-iterate its main observational features to then discuss possible implications under the one and the other hypothesis. 

The dip is a persistent and always well-defined structure, it does not show any energy dependency, and its width is the same in both observations. The ingress into and the egress from the dip are not resolved in our data. 
These properties clearly argue for the eclipse interpretation. In objects, that show both an eclipse and a stream-dip or just the dip, the phase and width of the dip is often remarkably variable, both in phase, showing a jitter or a trend with mass accretion rate, in its width, and its energy dependence \citep[see e.g.][for HU Aqr and EK UMa, respectively]{schwarz+09, clayton+osborne94}. Nothing of this applies to the feature in \xmm which is a strong argument for an eclipse but not finally excluding a dip.

If we assume that \xmm is eclipsing, the following constraints on its geometry can be derived. Assuming the secondary star has a mass of 0.14 $M_{\odot}$, following \cite{knigge+11}, and the white dwarf has a mass of 0.8 $M_{\odot}$, we infer a minimum inclination $i > 80.5\degr$ from the length of the eclipse. The center of the bright phase is located at $\phi_{\rm obs1} = 0.88\pm0.03$ and $\phi_{\rm obs2} = 0.84\pm0.03$. These values were determined as midpoints between phases of half-light during the rise and fall to the main hump. The phase difference between the center of the bright phase and the eclipse gives the longitude of the accretion region i.e. 0.12 and 0.16 phase unit difference between the bright phase center and eclipse. The found angle of 50$\pm$18 degrees is typical for polars \citep{cropper88}. 

The accretion region's co-latitude ($\beta$) is difficult or impossible to determine with some precision. The bright phase length, $\Delta \phi > 0.5$, places the observer and accretion zone in the same hemisphere compared to the orbital plane (the 'northern' or upper hemisphere). The region's radial and longitudinal extent and potential contributions from another location on the white dwarf limit our conclusions about $\beta$. Polarimetric observations are necessary for distinctive results \citep[see examples][]{bailey+81,breinerd+85,piirola+87}.


Regarding \xmmn, the eclipsing scenario is perhaps questionable in one component. If the inclination is high and the accretion region is on the upper hemisphere, the matter that feeds the accretion region is crossing the line of sight. Other polars then show a stream dip. A prominent example is HU Aqr \citep{schwope+01}. The absence of such a dip might challenge our proposed accretion geometry as sketched above. However, there are counterexamples. V808 Aur is almost a twin of HU Aqr, in particular regarding its accretion geometry, but does not show X-ray dips, only optical dips in the very high state \citep{worpel+15, schwope+15}. Furthermore, our spectral analysis has revealed a high absorbing column depth, which could imply absorption in an extended accretion curtain instead of a more focused stream.

We now briefly discuss the alternative scenario, the dip-like feature is a stream-dip. We then infer $i> \beta$, i.e.~likely in the range $45\degr< i<75\degr$, where the maximum value of $i$ is given by the lack of an eclipse. If \xmm is not eclipsing, true phase zero is not known and could be determined only through phase-resolved spectroscopy tracing featured originating from the secondary star (some absorption lines or narrow emission lines from the irradiated front side). Otherwise one only may assume, that \xmm shows an average behavior. If so, the stream dip might be suspected to occur at phase 0.85. True phase zero would occur somewhere at the end of the X-ray bright phase, hence the main accretion column would then be situated at a considerable far distance from the secondary star ($\psi$ > 90\degr). Such an accretion geometry is rather unusual for polars and therefore regarded as a further argument against the stream-dip hypothesis.

\section{Discussion and conclusion \label{s:dis}}

This study presents the results acquired by the integration of a comprehensive X-ray analysis of \xmmn with data collected from several sky surveys and the subsequent SALT observation of the \xmm. Together with the obtained data, we identified this object as a polar-type magnetic CV. 

\xmm has two distinct XMM-Newton observations conducted over a time interval of one year. The light curves exhibit complete similarity, and the periods derived from the period analysis demonstrate congruence, yielding identical values. The system has a singular period that corresponds with the synchronous rotation characteristic observed in polars. In particular, the power spectrum of Intermediate Polars' optical or X-ray light curves indicates additional frequencies. These frequencies occur due to the lack of synchronization between the orbital motion and the spin of the white dwarf. Periodograms often exhibit complex frequency distributions \citep{norton+96}. The distinct presence of a single period and its harmonics in the power spectrum, along with its short orbital period resembling polars \citep{ritter+03}, led to \xmm being classified as a polar. 

The X-ray light curves have a prominent hump-like structure that encompasses a distinct dip-like feature. The observed deep structure exhibits consistent widths, despite being derived from distinct levels of X-ray emissions. This feature has led to the assumption that it is an eclipse. The length of the deep eclipse-like feature is similar to those found for known eclipsing polars HU Aqr \citep{schwarz+09} and V808 Aur \citep{worpel+15}. In Obs 1, the HR variation is likely the same, changes between 0.0 - 0.2 in the bright phase, and is compatible with these known objects.

The low-resolution spectrum acquired from SALT observation has prominent emission lines corresponding to hydrogen (H) and helium (He), providing as the most conspicuous characteristic indicative of its polar nature. The prevalence of the He II line is a characteristic linked to magnetic cataclysmic variables \citep{szkody+02}. The spectrum displays two weak humps reminiscent of cyclotron humps observed in polars. The object is considered to be a white dwarf with a weak magnetic field.

By the above, it is evident from the almost simultaneous observations conducted by ZTF and ATLAS that the \xmm exhibits wide brightness variation over the years. This behavior has also been detected by the \ero All Sky Survey. This high variability is a phenomenon observed in magnetic CVs with variation in the amount of the mass flow and is common in some known polar \citep[AM Her,][]{wu+08} and intermediate polars \citep[MU Cam,][]{staude+08}. Nevertheless, it is important to highlight that the significant variation in brightness observed in the \xmm example might be attributed to observations obtained during distinct orbital phases, which exhibit pronounced eclipses. On the other hand, even in these distinct scenarios, the photometric period, X-ray behavior, and optical spectrum features of the system primarily indicate its polar nature.

Both \xmmn observations contain different X-ray levels. In essence, the temperatures we obtained are within the ranges indicated, especially for polars \citep{kuulkers+10}. The spectra obtained for Obs 1 exhibit a dominant presence of hard X-ray photons, which one can attribute to reflection. According to \cite{mukai17}, it has been observed that the temperature in the X-ray spectrum, particularly in magnetic CVs where hard X-rays are prevalent, might potentially be attributed to the phenomenon of reflection. The potential challenge described in the X-ray spectrum part can be effectively addressed by the implementation of a complex absorber, particularly in our case of the Obs 1. Nevertheless, this issue was not observed in Obs 2.

The X-ray spectra acquired from both observations exhibit an absence of the typical blackbody emission characteristic of polars. This feature was also not visible, particularly in the discovered polars by \xmmn \citep{ramsay+04} and first discovered polar by \ero All Sky Survey \citep{schwope+22}. Another intriguing finding derived from the spectra of the object is the high column densities derived from the bright phase, which is considerably higher than the galactic one. The significant column density can be attributed to either partial absorption by dense material or absorption by ionized material. The spectral fit in magnetic CVs frequently suggests the existence of a partial obscuring absorber, where some photons are detected directly while others can be seen through an intervening absorber. The 45\% partial absorption seen in Obs 1 appears to be in line with this. This physical behavior indicates that the absorber's size is comparable to the X-ray emission region, may be located close to the white dwarf or within the binary itself \citep{mukai17}. In the event that the deep dip corresponds to an eclipse, one can expect to observe the accretion stream 0.1-0.2 phases before deep dip in the bright phase. While the light curves do not provide a clear indication of the absorption impact caused by the accretion stream, the presence of a high column density in the spectrum suggests a potential correlation with this accretion element. In particular, the high column density obtained in the bright phase is similar to those previously obtained from EXOSAT observations of EF Eri, a known polar \citep{watson+89}. Normally, for EF Eri, these column densities were obtained at the dip regions related to the stream. In the case of \xmm, the source of this absorption is perhaps due to an extended accretion curtain covering the entire bright phase and covering the accretion column.

According to the previous results, we strongly suggest that \xmm is categorized as a polar-type CV. In addition to the data provided here, we have not identified any further information on the object in other data archives. The data employed in this study does not provide sufficient information to accurately describe the status of the eclipse-like feature. High-speed photometric measurements are necessary to have a more comprehensive understanding of this behavior. The addition of new samples to CVs is of great importance as they are crucial for understanding how these systems evolved and occurred as they are at the last step of stellar evolution, as well as understanding the distribution of magnetic CVs and the presence of magnetism in the Milky Way. Advanced tools like \ero now have the resolve power to detect these typically faint sources. 


\begin{acknowledgements}
This research has made use of data, software and/or web tools obtained from the High Energy Astrophysics Science Archive Research Center (HEASARC), a service of the Astrophysics Science Division at NASA/GSFC and of the Smithsonian Astrophysical Observatory’s High Energy Astrophysics Division. This work is based on data from eROSITA, the soft X-ray instrument aboard SRG, a joint Russian-German science mission supported by the Russian Space Agency (Roskosmos), in the interests of the Russian Academy of Sciences represented by its Space Research Institute (IKI), and the Deutsches Zentrum für Luft- und Raumfahrt (DLR). The SRG spacecraft was built by Lavochkin Association (NPOL) and its subcontractors, and is operated by NPOL with support from the Max Planck Institute for Extraterrestrial Physics (MPE). The development and construction of the eROSITA X-ray instrument was led by MPE, with contributions from the Dr. Karl Remeis Observatory Bamberg \& ECAP (FAU Erlangen-Nuernberg), the University of Hamburg Observatory, the Leibniz Institute for Astrophysics Potsdam (AIP), and the Institute for Astronomy and Astrophysics of the University of Tübingen, with the support of DLR and the Max Planck Society. The Argelander Institute for Astronomy of the University of Bonn and the Ludwig Maximilians Universität Munich also participated in the science preparation for eROSITA. The eROSITA data shown here were processed using the eSASS software system developed by the German eROSITA consortium. We thank the anonymous referee for useful comments and suggestions which helped to improve and clarify the paper.

\end{acknowledgements}

%
%

\bibliography{xmm152736}
\bibliographystyle{aa}

\end{document}